# The Science Case for Spacecraft Exploration of the Uranian Satellites

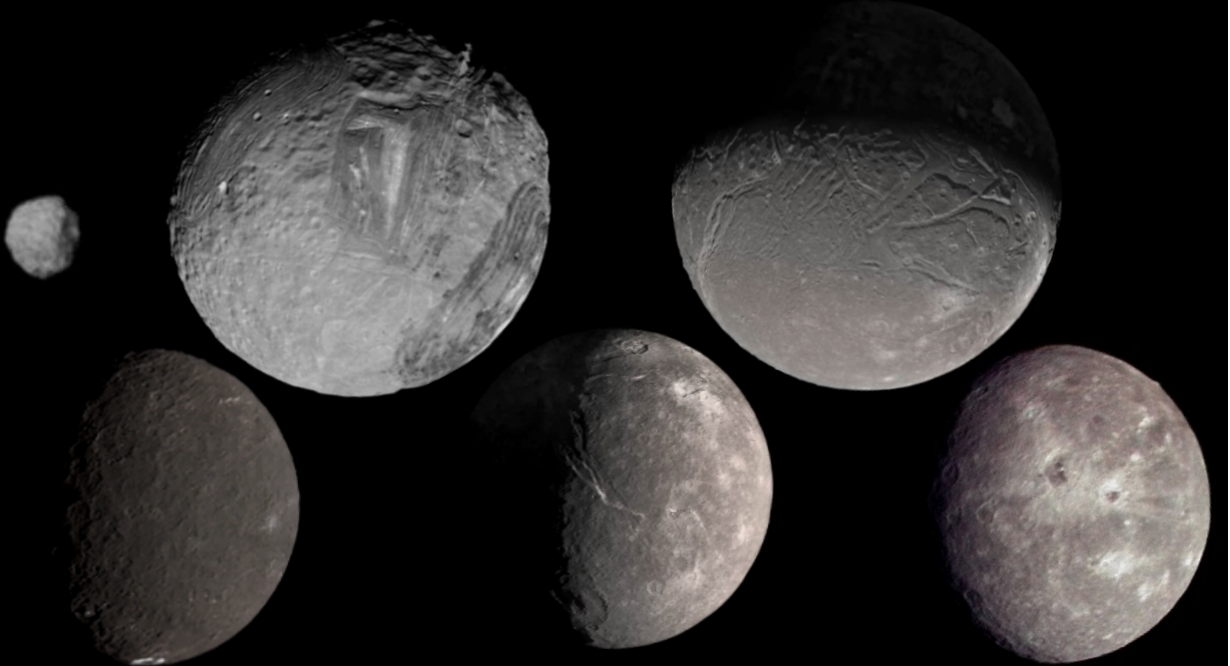

**Lead Authors:**
Richard J. Cartwright[1] & Chloe B. Beddingfield[1,2]
[1]SETI Institute, [2]NASA Ames Research Center

rcartwright@seti.org
chloe.b.beddingfield@nasa.gov

*Uranian satellites imaged by Voyager 2, not shown to scale (NASA/JPL-Caltech/USGS, [55]). Puck (top left), Miranda (top middle), Ariel (top right), Umbriel (bottom left), Titania (bottom middle), and Oberon (bottom right).*

**Co-Authors:**

| | |
|---|---|
| T. Nordheim, Jet Propulsion Laboratory | D. Burr, Northern Arizona University |
| C. Elder, Jet Propulsion Laboratory | A. Ermakov, University of CA Berkeley |
| W. Grundy, Lowell Observatory | J. Roser, SETI Institute & NASA Ames |
| B. Buratti, Jet Propulsion Laboratory | J. Castillo-Rogez, Jet Propulsion Laboratory |
| A. Bramson, Purdue University | M. Showalter, SETI Institute |
| M. Sori, Purdue University | I. Cohen, John Hopkins University, APL |
| R. Pappalardo, Jet Propulsion Laboratory | E. Turtle, John Hopkins University, APL |
| M. Neveu, Goddard Space Flight Center | M. Hofstadter, Jet Propulsion Laboratory |

**Additional Co-Authors and Endorsers:**

A full list of co-authors and endorsers is included at the end of this document.

# 1. Introduction and Motivation

> A spacecraft mission to the Uranian satellites would address the following 'Big Questions' identified in the *Scientific Goals for Exploration of the Outer Solar System* document, outlined by the OPAG community (https://www.lpi.usra.edu/opag/goals-08-28-19.pdf) [*Table 1*]:
>
> (**1**) What is the distribution and history of life in the Solar System?
> (**2**) What is the origin, evolution, and structure of planetary systems?
> (**3**) What present-day processes shape planetary systems, and how do these processes create diverse outcomes within and across different worlds?

The large moons of Uranus are possible ocean worlds [1] that exhibit a variety of surface features, hinting at endogenic geologic activity in the recent past (*e.g.*, [2]). These moons are rich in water ice, as well as carbon-bearing and likely nitrogen-bearing constituents, which represent some of the key components for life as we know it. However, our understanding of Uranus and its moons is severely limited by the absence of data collected by an orbiting spacecraft. **We assert that multiple close proximity flybys of the Uranian moons made by a Flagship-class spacecraft in orbit around Uranus is needed to conduct essential Solar System science, and initiation and design of this mission must occur in the upcoming decade (2023 – 2032)**.

An orbiter would vastly improve our understanding of these possible ocean worlds and allow us to assess the nature of water and organics in the Uranian system, thereby improving our knowledge of these moons' astrobiological potential. A Flagship mission to Uranus can be carried out with existing chemical propulsion technology by making use of a Jupiter gravity assist in the 2030 – 2034 timeframe, leading to a flight time of only ~11 years, arriving in the early to mid 2040's (outlined in the Ice Giants Pre-Decadal Survey Mission Study Report: https://www.lpi.usra.edu/icegiants/mission_study/Full-Report.pdf). Crucially, this arrival timeframe would allow us to observe the Uranian moons' northern hemispheres, which were shrouded by winter at the time of the Voyager 2 flyby and have never been imaged. An orbiter could then continually collect data and observe seasonal changes to the surfaces of these moons as the Uranian system transitions into southern spring in 2049. A complementary assessment of the science that could be achieved by a Flagship mission to the Uranus system is described in another paper submitted to the Planetary Science and Astrobiology Decadal Survey [3].

The five large moons of Uranus are enigmatic, with surfaces rich in volatiles and marked by bizarre landforms, hinting at geologically complex and recent activity. In 1986, the Voyager 2 spacecraft flew by the Uranian system and collected tantalizing snapshots of these 'classical' satellites, measured Uranus' offset and tilted magnetic field, as well as discovering ten new ring moons (*e.g.*, [2]). Since this brief flyby, investigation of Uranus and its rings and satellites has remained in the purview of ground- and space-based telescopes. Although these telescope observations have made some fascinating discoveries, many key science questions remain unanswered [*Table 1*]. Addressing these questions is vital for a fuller understanding of the Uranian system, which represents the highest unaddressed priority item from the last Planetary Science Decadal Survey (2013 – 2022). New measurements made by modern instruments on board an orbiting spacecraft are critical to investigate the surfaces and interiors of the large moons and determine whether they are ocean worlds with subsurface liquid water layers. Furthermore, a spacecraft mission to Uranus would enable a more complete investigation of organics and water in the outer Solar System, two of the key components for life as we know it, as well as improve our understanding of how geologic processes operate in cold and distant ice giant systems.

Table 1: **Science questions requiring exploration by a Flagship-class Uranus orbiter.**

| SCIENCE QUESTIONS | MEASUREMENTS | INSTRUMENTS |
|---|---|---|
| Do the satellites have subsurface oceans that are, or were, harbors for life? Are there signs of communication between their surfaces and interiors? Are any of these moons geologically active? **Addresses OPAG Big Questions #1, 2, 3** | Search for induced magnetic fields, plumes, hot spots, cryovolcanic features, and surface changes since the Voyager 2 flyby, and search for dust samples from possible plume sources | Magnetometer VIS camera Mid-IR camera VIS/NIR mapping spectrometer Dust spectrometer |
| What are the internal structures of the classical satellites? **Addresses OPAG Big Questions #2, 3** | Moment of inertia measurements, gravity field characterization, analysis of geologic, topographic, spectral maps, and magnetic induction | Radio science subsystem VIS camera VIS/NIR mapping spectrometer Magnetometer |
| What processes modify the satellites and what are the compositions of their geologic units and features? **Addresses OPAG Big Questions #2, 3** | Analysis of geologic, topographic, and spectral maps, estimate surface ages from impact crater densities | VIS camera VIS/NIR mapping spectrometer |
| Do the moons have tenuous atmospheres? Do volatiles migrate seasonally? **Addresses OPAG Big Questions #2, 3** | Search for exospheres and changes in the distribution and spectral signature of condensed volatiles | VIS camera VIS/NIR mapping spectrometer Plasma spectrometer |
| Do magnetospheric charged particles weather the surfaces of the ring moons and classical satellites? **Addresses OPAG Big Questions #2, 3** | Characterize magnetic field and charged particle populations proximal to moons | Magnetometer Plasma spectrometer Energetic particle detector |
| Is the red material on the classical satellites organic-rich and did it originate from the irregular satellites? **Addresses OPAG Big Questions #2, 3** | Spectral maps of classical moons, inbound flyby of an irregular satellite, images of the irregular moons while in Uranus orbit, and collect and analyze dust samples | VIS camera VIS/NIR mapping spectrometer Dust spectrometer |
| Does Mab sustain the $\mu$-ring? Does $\mu$-ring material coat other moons? **Addresses OPAG Big Questions #2, 3** | Spectral mapping of the ring moons and Miranda, and collect and analyze $\mu$-ring dust samples | VIS camera VIS/NIR mapping spectrometer Dust spectrometer |
| What is the dynamic history of the moons? Were there previous orbital resonances? **Addresses OPAG Big Questions #2, 3** | Eccentricities, inclinations, tidal $Q(\omega)$ of Uranus, Love numbers, paleo heat fluxes | Radio science subsystem VIS Camera |

## 2. Geology of the Uranian Satellites

Classical Moons: Voyager 2 collected fascinating images of the five large moons' southern hemispheres (subsolar point ~81ºS) [*Figure 1*], but their northern hemispheres were shrouded by winter darkness at the time of the flyby. The incomplete spatial coverage, and generally low spatial resolution of the available images, limits our understanding of different terrains and geologic features, in particular for the more distant moons Umbriel, Titania, and Oberon.

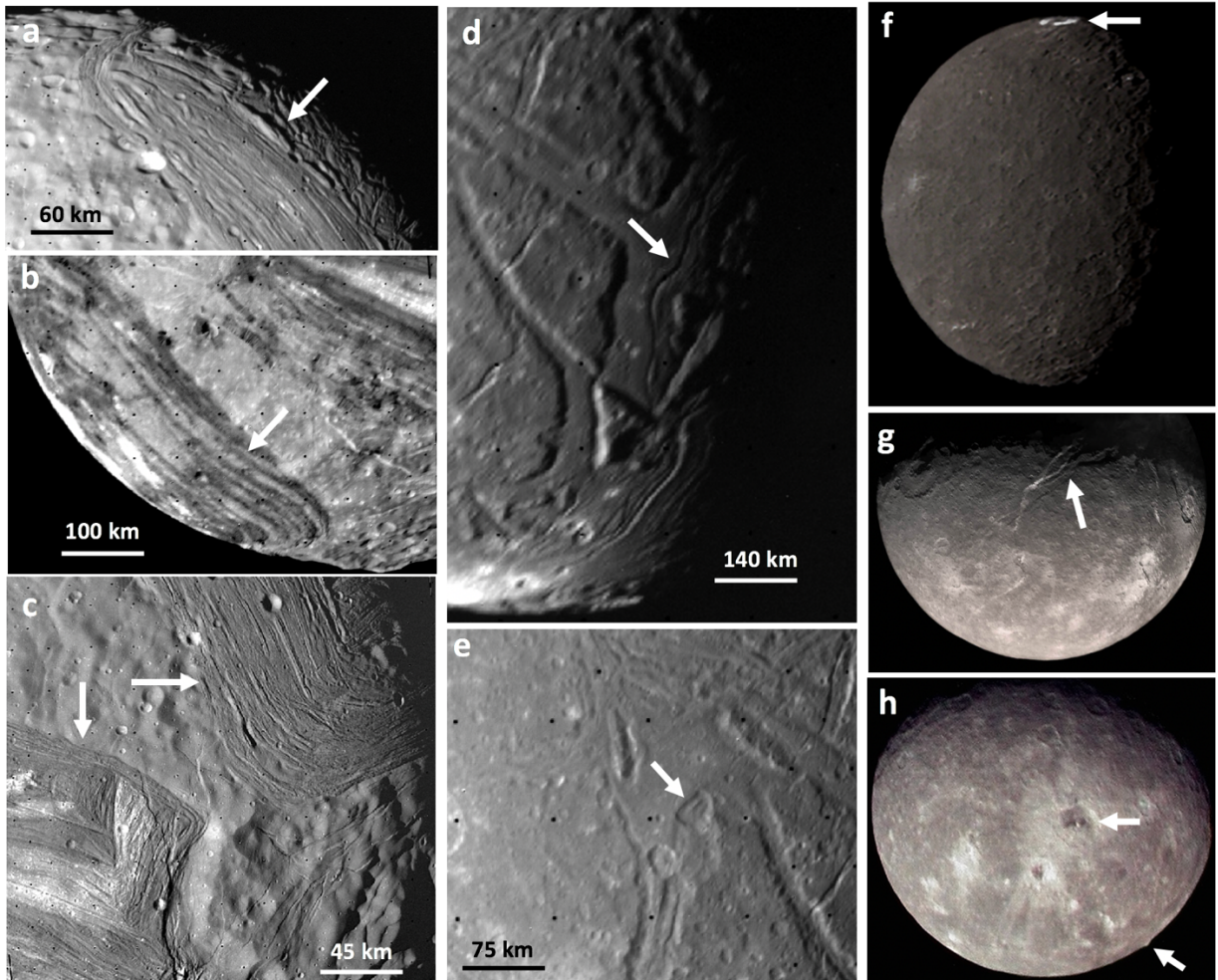

*Figure 1:* *Voyager 2 images of the Uranian moons. White arrows highlight: (a) ridges on Miranda, which possibly have a cryovolcanic and tectonic origin; (b) Arden Corona on Miranda with high and low albedo banding along large tectonic faults; (c) Inverness (bottom left) and Elsinore (top right) Coronae on Miranda that exhibit ridges and grooves. Between these two coronae are examples of craters that have been mantled by an unknown source of regolith; (d) large chasmata with medial grooves on Ariel; (e) an impact crater on Ariel, possibly infilled by cryolava; (f) the bright floor of Wunda crater on Umbriel; (g) the large Messina Chasmata on Titania; and (h) the smooth floor of Hamlet crater and an 11 km tall 'limb mountain' on Oberon.*

The innermost moon Miranda displays abundant evidence for endogenic geologic activity, including three large polygonal shaped regions called coronae, which were likely formed by tectonic and/or cryovolcanic processes (*e.g.*, [2, 4-11]) [*Figure 1a-c*]. The origin and time scale of activity on Miranda is not well understood, and it is unknown if this activity is associated with a subsurface ocean, either now or in the past. Investigation of induced magnetic fields, plumes, surface heat anomalies, as well as analysis of geologic surface features interpreted to be cryovolcanic is paramount to determine if Miranda is an ocean world. Tidal heating of Miranda from past orbital resonances [12,13] may have been an important driver of resurfacing in the recent past. Additionally, Miranda displays ancient cratered terrain, pockmarked with 'subdued' craters, which have been mantled by an unknown source of material (*e.g.*, [2,11]). These craters are reminiscent of the plume-mantled craters on the ocean world Enceladus [11], hinting at a similar plume-driven mantling process on Miranda. Determining whether Miranda is, or was, an ocean world requires high resolution datasets and multiple close flybys, which can only be collected by an orbiter.

Miranda's neighbor Ariel also displays widespread evidence for resurfacing, with chasmata dominating large sections of its surface [*Figure 1d, 1e*]. The smooth floors of some of these chasmata are bowed up with two parallel medial ridges that are separated by a topographic low, reminiscent of fissure style volcanism on Earth (*e.g.*, [7]). Large fracture systems cut across other parts of Ariel's surface, and clusters of curvilinear features referred to as 'flow bands' are thought to be cryovolcanic features (*e.g.*, [5,7,14,15]). Much of Ariel's surface is relatively young (~1 – 2 Ga) [16], but the process(es) that resurfaced this moon are poorly understood [15,17]. Investigation of possible cryovolcanism and geologic communication between the interior and surface of Ariel requires high resolution data collected during multiple close flybys by an orbiter.

Although Umbriel has the darkest and oldest surface of the five classical moons (*e.g.*, [16,18]), it exhibits evidence for global-scale resurfacing [18] and large craters like Wunda, which has a bright annulus of material surrounding its central peak (*e.g.*, [2]) [*Figure 1f*]. This bright annulus may represent a large deposit of $CO_2$ ice [19-21] that originated from post-impact cryovolcanic infilling (*e.g.*, [2,5]). However, the resolution of the available data limits analyses of these features, and new images collected during close flybys are needed to understand Umbriel's geologic history.

The surfaces of the outer moons Titania and Oberon exhibit evidence for tectonism, with large chasmata and linear surface features, as well as smooth plains units that may have resulted from cryovolcanic processes (*e.g.*, [2,4]) [*Figure 1g, 1h*]. Additionally, Oberon has an ~11 km tall 'limb mountain' that could be the central peak for a relaxed complex crater (*e.g.*, [2, 22]). Geologic analyses of these tectonic and cryovolcanic features, and Oberon's curious limb mountain, are not possible without higher spatial resolution datasets collected by an orbiter.

Internal Structure: Investigating the internal structure and bulk composition of the classical moons is critical for understanding the formation and evolution of the Uranian satellites. The densities of Ariel, Umbriel, Titania, and Oberon range from 1.5 to 1.7 g cm$^{-3}$, indicating that these moons are made of at least 50% silicate material by mass, whereas Miranda's density could be as low as 1.2 g cm$^{-3}$, indicating a larger $H_2O$ ice fraction. Measuring the long-wavelength shape [23, 24], and non-spherical gravity field would shed light on the differentiation state and the nature of the endogenic activity exhibited by these moons, as was done by Galileo and Cassini for the Jovian and Saturnian moons. The existence of subsurface oceans could be revealed by measuring libration amplitudes [25], as well as by measuring magnetic induction with a magnetometer. Furthermore, mapping heat fluxes across the surfaces of these moons, using a mid-IR camera, is important for understanding their heat budgets and the long-term survivability of liquid water in their interiors. These measurements can only be made by an orbiter making multiple close passes of the moons.

Ring Moons and Irregular Satellites: Voyager 2 initially discovered ten ring moons: Cordelia, Ophelia, Bianca, Cressida, Desdemona, Juliet, Portia, Rosalind, Belinda, and Puck (*e.g.*, [2]). Perdita was discovered later through reanalysis of Voyager 2 images [26]. Cupid and Mab were discovered using the Hubble Space Telescope [27,28]. Mab orbits within the outermost and dusty $\mu$-ring, which might be sustained by material ejected from the surface of this tiny moon [27,28]. Furthermore, $\mu$-ring material might mantle other moons including Miranda, possibly contributing to its substantial regolith cover [29]. Little is known about the surface geologies of the ring moons as only Puck was spatially resolved by Voyager 2. The Voyager 2 images revealed a heavily cratered surface, suggesting that Puck may have been collisionally disrupted and then reaccreted into a rubble pile [2,5]. Voyager 2 did not detect any of Uranus' nine known irregular satellites, which were discovered by ground-based observers (*e.g.*, [30-32]). Thus, the geologies of Uranus' ring moons and irregular satellites remain unexplored, and new observations made by an orbiter would dramatically expand our knowledge of these objects. A complementary assessment of the ring moon and irregular satellite science that could be achieved by a Uranus orbiter is described in another paper submitted to the Planetary Science and Astrobiology Decadal Survey [33].

## 3. Surface Compositions of the Uranian Satellites

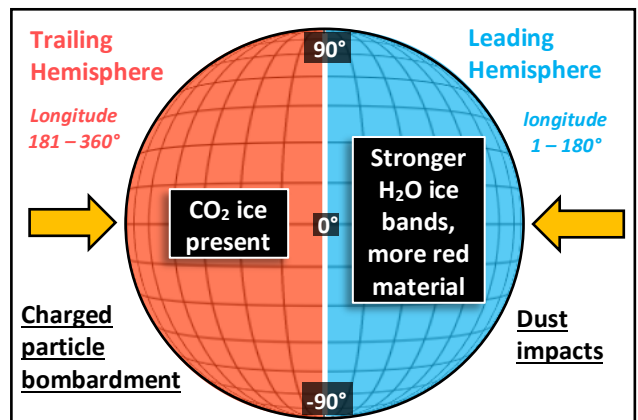

*Figure 2: Illustration showing the broad leading/trailing trends in composition exhibited by Ariel, Umbriel, Titania, and Oberon, possibly driven by charge particle interactions (primarily trailing) and dust impacts (primarily leading).*

Classical Moons: Ground-based telescope observations determined that the classical moons have surface compositions dominated by a mixture of $H_2O$ ice and a dark, spectrally neutral material of unknown origin (*e.g.*, [34-36]). Laboratory experiments indicate that the dark material has a spectral signature similar to amorphous carbon and/or silicates [37]. More recent observations have determined that the Uranian satellites display leading/trailing and planetocentric asymmetries in their compositions [**Figure 2**]. For example, 'pure' $CO_2$ ice (*i.e.*, segregated from other constituents in concentrated deposits) has been detected, primarily on the trailing hemispheres of the inner moons, Ariel and Umbriel [19,20,38,39]. $CO_2$ ice on these moons could be generated via irradiation of native $H_2O$ ice and C-rich material by magnetospheric charged particles [19,20]. Furthermore, $H_2O$ ice bands are weaker on the trailing hemispheres of these moons, perhaps in part due to large deposits of $CO_2$ ice masking $H_2O$ [19,20]. Another possibility is that heliocentric dust impacts promote regolith overturn and expose 'fresh' $H_2O$ ice, primarily on these moons' leading hemispheres [39]. Spectrally red material has also been detected, primarily on the leading hemispheres of the outer moons, Titania and Oberon (e.g., [40]). The distribution of red material could result from the accumulation of infalling dust from retrograde irregular satellites [39-41], which are spectrally redder than the classical moons (*e.g.*, [39,42,43]).

At longer wavelengths (~3 – 5 μm), the spectral signature of 'pure' $CO_2$ ice is strangely absent from these moons [20,29,39]. One possible explanation is that the classical moons have regoliths mantled by a thin layer of tiny $H_2O$ ice grains (≤ 2 μm diameters), which enhance surface scattering and obscure larger grains of $CO_2$ retained beneath this topmost layer [20,29,39]. Supporting this possibility, visible wavelength polarimetry data suggest that these moons have porous regoliths, dominated by tiny ice grains [44]. These datasets indicate that the regoliths of the Uranian moons are starkly different from both $H_2O$ ice-rich and dark material-rich Galilean and Saturnian moons.

Although these hypotheses are intriguing, new spacecraft measurements are needed to identify the processes modifying the surface compositions of the Uranian moons. For example, the distribution of $CO_2$ ice is only longitudinally constrained, limiting our ability to determine whether this volatile is generated by charged particle radiolysis, or whether it is a native constituent sourced from their interiors. Unlike the other classical satellites, $CO_2$ ice and hemispherical asymmetries in composition are absent on Miranda [19,20,30,39,45], adding to the mystery surrounding this moon. Some ground-based spectra suggest that $NH_3$-bearing species are present [39,45,46], which are highly efficient anti-freeze agents that could promote the retention of subsurface oceans if present within the interiors of these moons [11,15]. Higher spatial resolution spectra are needed to determine whether this $NH_3$-rich material is spatially associated with geologic features, suggesting an endogenic origin, as seen in spacecraft data for other icy bodies like Charon (*e.g.*, [47]). The regolith properties of the Uranian satellites could result from interactions with the surrounding space environment, which cannot be properly assessed without data collected by an orbiter.

Furthermore, the composition and origin of the red material and the ubiquitous dark material on the Uranian moons remains poorly understood. These materials could represent organic-rich constituents delivered to and/or native to these moons. Prior spacecraft missions have assessed the nature of organics in the Jupiter, Saturn, and Pluto systems, as well as on comets. Investigating organics in the Uranian system represents a key heliocentric link for improving our understanding of the nature and overall distribution of organic matter in the Solar System, as well as for investigating whether organics formed within the protoplanetary disk or were delivered as interstellar matter. Thus, new measurements made by an orbiter are critical for determining the spatial distribution and spectral signature of volatile constituents on these moons and for investigating the origin and evolution of organic material in the Solar System.

Ring Moons and Irregular Satellites: Far less is known about the compositions of Uranus' smaller ring moons and irregular satellites, which are too faint for spectroscopic observations using available telescopes. Photometric datasets indicate that the ring moons are dark, with neutral spectral slopes and slight reductions in albedo at 1.5 and 2.0 μm, hinting at the presence of $H_2O$ ice (*e.g.*, [48]). Other photometric studies determined that Uranus' irregular satellites are dark and red (*e.g.*, [42,43]), with redder colors than the irregular satellites of the other giant planets [49]. Spectra of the largest Uranian irregular satellite Sycorax suggest that $H_2O$ ice is present [50], but no spectra exist for the other, fainter moons. Therefore, the compositions of these objects are essentially unknown, and new observations made by an orbiter are needed.

## 4. Conclusions and Recommendations for Future Exploration

Data returned by Voyager 2 and ground- and space-based telescopes have revealed tantalizing glimpses of the Uranian moons' geologic histories, compositions, and interactions with the surrounding space environment. However, the flyby nature of the Voyager 2 encounter, the lack of a mapping spectrometer, and the low spatial resolution of collected datasets left many unanswered questions. Future telescope facilities like the Extremely Large Telescopes (ELTs) and the proposed Large UV/Optical/IR Surveyor (LUVOIR) space telescope will be able to collect high quality images and spectra of the classical satellites [51,52], providing new information about their surfaces [***Figure 3***]. Although these telescope datasets will undoubtedly increase our knowledge, they will not be able to assess the astrobiological potential of these possible ocean worlds, the linkages between geologic features and their surface compositions, nor probe their internal structures, or investigate moon-magnetosphere interactions [53], and they will not be able to resolve the ring moons and irregular satellites to assess their surface geologies and origins. **We recommend that a Flagship-class mission to the Uranus system be made a priority by NASA for the upcoming decade in order to address these critical topics**.

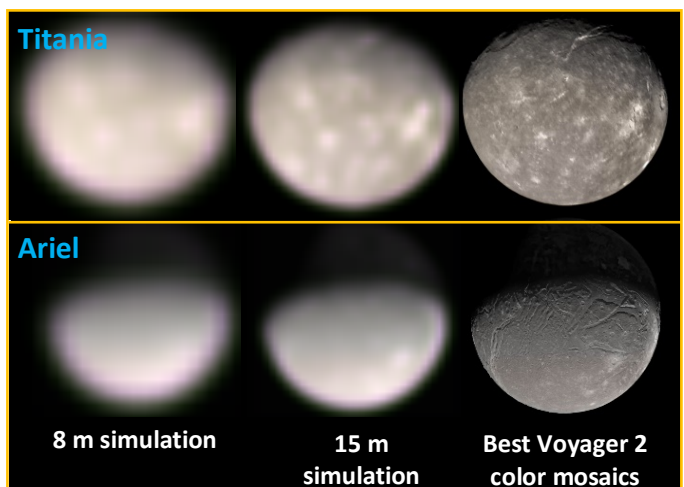

*Figure 3:* Resampled and real images of Titania and Ariel. Real images (right) are Voyager 2 image mosaics. Resampled images simulate what these moons would look like as seen by LUVOIR with an 8 m (left) and 15 m (center) aperture.

An orbiting spacecraft equipped with a carefully considered instrument suite [*Table 1*] could search for plumes and other signs of recent endogenic geologic activity. Furthermore, an orbiter would dramatically improve our understanding of the ring moons and could investigate whether Mab is the source of the $\mu$-ring by making multiple close passes of Uranus' rings. An orbiter could spend time looking outward, making key observations of the distant irregular satellites, similar to Cassini's observations of Saturn's irregular moons [54]. A close pass of an irregular satellite inbound to Uranus, like Cassini's inbound flyby of Phoebe, would represent an unparalleled opportunity to investigate the nature and origin of these likely captured objects. Thus, the moons of Uranus remain poorly understood, and new datasets returned by an orbiter are essential to peel back the veil shrouding the icy residents of the Uranian system.

**Full List of Co-Authors and Endorsers:**

| | |
|---|---|
| T. Nordheim | Jet Propulsion Laboratory |
| C. Elder | Jet Propulsion Laboratory |
| W. Grundy | Lowell Observatory |
| B. Buratti | Jet Propulsion Laboratory |
| A. Bramson | Purdue University |
| M. Sori | Purdue University |
| R. Pappalardo | Jet Propulsion Laboratory |
| M. Neveu | Goddard Space Flight Center |
| D. Burr | Northern Arizona University |
| A. Ermakov | University of California Berkeley |
| J. Roser | SETI Institute & NASA Ames Research Center |
| J. Castillo-Rogez | Jet Propulsion Laboratory |
| M. Showalter | SETI Institute |
| I. Cohen | John Hopkins University, Applied Physics Laboratory |
| E. Turtle | John Hopkins University, Applied Physics Laboratory |
| M. Hofstadter | Jet Propulsion Laboratory |
| E. Leonard | Jet Propulsion Laboratory |
| I. de Pater | University of California Berkeley |
| D.A. Patthoff | Planetary Science Institute & Jet Propulsion Laboratory |
| A. Masters | Imperial College London |
| L. Fletcher | University of Leicester |
| C. Ahrens | University of Arkansas |
| C. Andres | University of Western Ontario |
| K. Aplin | University of Bristol |
| G. Arney | NASA Goddard Space Flight Center |
| K. Baillié | Paris Observatory at Meudon |
| E. Barth | Southwest Research Institute |
| C. Bennett | University of Central Florida |
| R. Beyer | SETI Institute & Nasa Ames Research Center |
| C. Bierson | University of California Santa Cruz |
| M. Bland | US Geological Survey, Astrogeology Science Center |
| V. Bray | Lunar and Planetary Laboratory |
| P. Byrne | North Carolina State University |
| N. Cabrol | SETI Institute |
| M. Cameron | Jet Propulsion Laboratory |
| N. Chanover | New Mexico State University |
| C. Cochrane | Jet Propulsion Laboratory |
| G. Collins | Wheaton College |
| J. Cook | Pinhead Institute |
| A. Coustenis | Paris Observatory at Meudon |
| D. Cruikshank | NASA Ames Research Center |
| M. Ćuk | SETI Institute |
| I. Daubar | Jet Propulsion Laboratory & Brown University |
| A. Denton | Purdue University |
| D. DeColibus | New Mexico State University |
| R. Dhingra | University of Idaho |
| C. Dong | Princeton University |
| S. Ferguson | Arizona State University |
| G. Filacchione | INAF - Institute for Space Astrophysics and Planetology |
| R. French | SETI Institute |
| K. Golder | University of Tennessee |
| C. Grava | Southwest Research Institute |
| L. Griton | Paris Observatory at Meudon |
| N. Hammond | Wheaton College |
| A. Hayes | Cornell University |
| E. Hawkins | Loyola Marymount University |
| P. Helfenstein | Planetary Science Institute |
| A. Hendrix | Planetary Science Institute |
| A. Hofmann | Jet Propulsion Laboratory |
| B. Holler | Space Telescope Science Institute |
| T. Holt | University of Southern Queensland |
| S. Howell | Jet Propulsion Laboratory |
| C. Howett | Southwest Research Institute |
| H. Hussmann | Institute of Planetary Research |
| H. Hsu | University of Colorado |
| N. Izenberg | Johns Hopkins University Applied Physics Laboratory |
| R. Jacobsen | University of Tennessee |
| D. Jha | MVJ College of Engineering |
| R. Juanola-Parramon | NASA Goddard Space Flight Center |
| I. Jun | Jet Propulsion Laboratory |
| J. Keane | Jet Propulsion Laboratory |
| E. Karkoschka | Lunar and Planetary Laboratory |
| S. Kattenhorn | University of Alaska |
| M. Kinczyk | North Carolina State University |
| M. Kirchoff | Southwest Research Institute |
| C. Klimczak | University of Georgia |
| P. Kollmann | Johns Hopkins University Applied Physics Laboratory |
| R. Lopes | Jet Propulsion Laboratory |
| M. Lucas | University of Tennessee |
| A. Lucchetti | Astronomical Observatory of Padova |
| E. Martin | Smithsonian Institute |
| S. MacKenzie | Johns Hopkins University Applied Physics Laboratory |
| J. Moses | Space Science Institute |
| A. Barr Mlinar | Planetary Science Institute |
| J. Moore | NASA Ames Research Center |
| F. Nimmo | University of California Santa Cruz |
| S. O'Hara | Lunar and Planetary Institute |
| M. Pajola | Astronomical Observatory of Padova |
| S. Peel | University of Tennessee |
| G. Peterson | University of British Columbia |
| N. Pinilla-Alonso | University of Central Florida, Florida Space Institute |
1

| | |
|---|---|
| S. Porter | Southwest Research Institute |
| F. Postberg | Free University of Berlin |
| M. Poston | Southwest Research Institute |
| A. Probst | Jet Propulsion Laboratory |
| S. Protopapa | Southwest Research Institute |
| L. Quick | NASA Goddard Space Flight Center |
| A. Ricca | SETI Institute & NASA Ames Research Center |
| A. Roberge | NASA Goddard Space Flight Center |
| J. Roberts | Johns Hopkins University Applied Physics Laboratory |
| S. Robbins | Southwest Research Institute |
| S. Rodriguez | Université Paris Diderot |
| K. Runyon | Johns Hopkins University Applied Physics Laboratory |
| P. Schenk | Lunar and Planetary Institute |
| M. Schneegurt | Wichita State University |
| F. Scipioni | SETI Institute |
| M. Shusterman | Arizona State University |
| K. Singer | Southwest Research Institute |
| K. Soderlund | University of Texas |
| J. Spencer | Southwest Research Institute |
| L. Spilker | Jet Propulsion Laboratory |
| K. Stephan | German Aerospace Center Institute |
| T. Stryk | Roane State Community College |
| M. Tiscareno | SETI Institute |
| T. Tomlinson | University of Colorado |
| F. Tosi | INAF - Institute for Space Astrophysics and Planetology |
| P. Tortora | Alma Mater Studiorum - Università di Bologna |
| E. Turtle | Johns Hopkins University Applied Physics Laboratory |
| O. Umurhan | SETI Institute & NASA Ames Research Center |
| S. Vance | Jet Propulsion Laboratory |
| A. Verbiscer | University of Virginia |
| C. Walker | Woods Hole Oceanographic Institution |
| B. Weiss | Massachusetts Institute of Technology |
| O. White | SETI Institute & NASA Ames Research Center |
2